\documentclass[10pt,conference]{IEEEtran}

\usepackage[english]{babel} 
\usepackage{amsmath}                
\usepackage{amsfonts}               
\usepackage{amssymb}                
\usepackage{amsopn}                 
\allowdisplaybreaks[3] 
\usepackage{bbm}                    
\usepackage{mathrsfs}               
\usepackage{calc}                   
\usepackage[dvips]{graphicx}        
\usepackage{epsfig}                
\usepackage{psfrag}                 
\usepackage[dvips]{color}           
\usepackage{fancyhdr}               
\usepackage{verbatim}               
\usepackage{exscale}                
\usepackage{macros}
\newtheorem{theorem}{Theorem}
\newtheorem{lemma}[theorem]{Lemma}

\title{On Multipath Fading Channels at High SNR}

\author{\IEEEauthorblockN{Tobias Koch ~~~~~ Amos Lapidoth}
\IEEEauthorblockA{ETH
  Zurich, Switzerland\\
Email: \{tkoch, lapidoth\}@isi.ee.ethz.ch}}

\begin{document}

\maketitle

\begin{abstract}
  This paper studies the capacity of discrete-time multipath fading
  channels. It is assumed that the number of paths is finite, i.e.,
  that the channel output is influenced
  by the present and by the L previous channel inputs. A
  noncoherent channel model is considered where neither
  transmitter nor receiver are cognizant of the fading's realization,
  but both are aware of its statistic. The focus is on capacity at high
  signal-to-noise ratios (SNR). In particular, the capacity
  \emph{pre-loglog}---defined as the limiting ratio of the capacity to
  loglog(SNR) as SNR tends to infinity---is studied. It is shown
  that, irrespective of the number of paths L, the capacity
  pre-loglog is 1.
\end{abstract}

\section{Introduction}
\label{sec:intro}
This paper studies the capacity of multipath (frequency-selective) fading
channels. A noncoherent channel model is considered where neither
transmitter nor receiver are cognizant of the fading's realization,
but both are aware of its statistic. 
Our focus is on the high signal-to-noise ratio (SNR) regime. 

It has
been shown in \cite{lapidothmoser03_3} for noncoherent
\emph{frequency-flat} fading channels that if the fading process is
of finite entropy rate, then at high SNR capacity grows
double-logarithmically with the SNR.\footnote{It is well known that when
  the receiver knows the fading perfectly, 
then capacity increases logarithmically in the
SNR \cite{ericson70}. Thus communicating over noncoherent
flat-fading channels at high SNR is power inefficient.} For
noncoherent \emph{multipath} fading channels, it has been recently demonstrated
that if the delay spread is large in the sense that the
variances of the path gains do not decay faster than geometrically,
then capacity is \emph{bounded} in the SNR
\cite{kochlapidoth08_1}. For such channels, capacity does not tend
to infinity as the SNR tends to infinity.

The above condition can only be met by multipath fading channels that
have an \emph{infinite} number of paths in the sense that the channel output is
influenced by the present and by \emph{all} previous channel inputs. In this paper we
consider multipath fading channels with a \emph{finite} number of
paths, i.e., the channel output is only influenced by the present and by the $\const{L}$ previous
channel inputs. In order to characterize the capacity of this channel
at high SNR, we study the capacity \emph{pre-loglog}, defined as the limiting
ratio of capacity to $\log\log\SNR$ as SNR tends to infinity. We show
that the pre-loglog is not diminished by the multipath behavior, i.e.,
irrespective of the value of $\const{L}$ the pre-loglog is $1$. To state this
result precisely we begin with a mathematical description of the
channel model.

\subsection{Channel Model}
\label{sub:channel}
Let $\Complex$ and $\Integers^+$ denote the set of complex numbers and the
set of positive integers, respectively. We consider a discrete-time
multipath fading channel whose channel output $Y_k \in \Complex$ at
time $k \in \Integers^+$ corresponding to the time-$1$ through
time-$k$ channel inputs $x_1,\ldots,x_k\in\Complex$ is given by
\begin{IEEEeqnarray}{c}
  Y_k = \left\{
  \begin{array}{ll}
    \displaystyle \sum_{\ell=0}^{k-1}
    H_k^{(\ell)}x_{k-\ell}+Z_k, \quad & k=1,\ldots,\const{L}\\
    \displaystyle \sum_{\ell=0}^{\const{L}}
    H_k^{(\ell)}x_{k-\ell}+Z_k, & k=\const{L}+1,\const{L}+2,\ldots \,.
  \end{array}\right.\;\;\label{eq:channel}
\end{IEEEeqnarray}
Here $H_k^{(\ell)}$ denotes the time-$k$ gain of the $\ell$-th path;
$\{Z_k\}$ is a sequence of independent and identically
distributed (IID), zero-mean, variance-$\sigma^2$,
circularly-symmetric, complex Gaussian
random variables; and $\const{L}\in\Integers_0^+$ (where $\Integers_0^+$
denotes the set of nonnegative integers) denotes the number of
paths that influence $Y_k$. For $\const{L}=0$, the channel \eqref{eq:channel}
reduces to the flat-fading channel that was studied in
\cite{lapidothmoser03_3}; and for $\const{L}=\infty$, Equation \eqref{eq:channel}
describes the multipath fading channel that was studied in \cite{kochlapidoth08_1}. In this
paper we shall focus on the case where the number of paths is finite, i.e.,
where $\const{L}<\infty$. We assume that for each path $\ell=0,\ldots,\const{L}$
 the stochastic process $\big\{H_k^{(\ell)}, k\in\Integers^+\big\}$
 is a zero-mean stationary process. We denote its variance and its
 differential entropy rate by
\begin{equation}
  \alpha_{\ell} \triangleq \E{\big|H_k^{(\ell)}\big|^2}, \qquad \ell=0\ldots,\const{L}
\end{equation}
and
\begin{equation}
  h_{\ell} \triangleq
  \lim_{n\to\infty}\frac{1}{n}h\left(H_{1}^{(\ell)},\ldots,H_n^{(\ell)}\right),
  \quad \ell =0,\ldots,\const{L},
\end{equation}
respectively. Without loss of generality, we assume that $\alpha_0>0$. We further assume that
\begin{equation}
  \alpha_{\ell} <\infty, \qquad \ell=0,\ldots,\const{L}
\end{equation}
and
\begin{equation}
  \min_{\ell\in\set{L}} h_{\ell}
  > -\infty,\label{eq:entropyrate}
\end{equation}
where the set $\set{L}$ is defined as $\set{L}\triangleq
\{\nu=0,\ldots,\const{L}:\alpha_{\nu}>0\}$.
We finally assume that the $\const{L}+1$
processes 
\begin{equation*}
\big\{H_k^{(0)},
k\in\Integers^+\big\},\ldots, \big\{H_k^{(\const{L})},
k\in\Integers^+\big\}
\end{equation*}
are independent (``uncorrelated scattering''), that they are jointly
independent of $\{Z_k\}$, and that the joint law of
\begin{equation*}
  \left(\{Z_k\},\big\{H_k^{(0)},
k\in\Integers^+\big\},\ldots, \big\{H_k^{(\const{L})},
k\in\Integers^+\big\}\right)
\end{equation*}
does not depend on the input sequence $\{x_k\}$. We consider a
\emph{noncoherent} channel model where neither the transmitter nor
the receiver is cognizant of the realization of $\big\{H_k^{(\ell)},
k\in\Integers^+\big\}$, $\ell=0,\ldots,\const{L}$, but both are aware of their
law. We do not assume that the path gains are Gaussian.

\subsection{Channel Capacity}
\label{sub:capacity}
\const{L}et $A_m^n$ denote the sequence $A_m,\ldots,A_n$. We define the
\emph{capacity} as
\begin{equation}
  C(\SNR) \triangleq \lim_{n\to\infty} \frac{1}{n}\sup I\big(X_1^n;Y_1^n\big),\label{eq:capacity}
\end{equation}
where the supremum is over all joint distributions on
$X_1,\ldots,X_n$ satisfying the power constraint
\begin{equation}
  \frac{1}{n}\sum_{k=1}^n\E{|X_k|^2}\leq\const{P}, \label{eq:power}
\end{equation}
and where SNR is defined as
\begin{equation}
  \SNR \triangleq \frac{\const{P}}{\sigma^2}.
\end{equation}
It can be shown that for the above channel model
\begin{IEEEeqnarray}{lCl}
  \IEEEeqnarraymulticol{3}{l}{\sup I(X_1^m;Y_1^m) + \sup
    I(X_1^n;Y_1^n)} \nonumber\\
  \quad & \leq & \sup
  I(X_1^{m+n};Y_1^{m+n}) + o(m+n), \quad m,n \in \Integers^+ \IEEEeqnarraynumspace
\end{IEEEeqnarray}
(where $\lim_{(m+n)\to\infty} o(m+n)/(m+n)=0$) so that, by a trivial generalization of Fekete's
lemma\footnote{Fekete's lemma states that if a sequence $\{a_n\}$ is
  \emph{superadditive}, i.e., $a_n+a_m\leq a_{m+n}$,
  $m,n\in\Integers^+$, then the limit $\lim_{n\to\infty}a_n/n$ exists
  and is given by $\sup_{n\in\Integers^+} a_n/n$.}, the limit in
\eqref{eq:capacity} exists and is given by
\begin{equation}
  \lim_{n\to\infty}\frac{1}{n}\sup I\big(X_1^n;Y_1^n\big) =
  \sup_{n\in\Integers^+} \frac{1}{n}\sup I\big(X_1^n;Y_1^n\big).
\end{equation}

By Fano's inequality, no rate above $C(\SNR)$ is achievable. (See
\cite{coverthomas91} for a definition of an achievable rate.) We do
not claim that there is a coding theorem associated with
\eqref{eq:capacity}, i.e., that $C(\SNR)$ is achievable. A coding
theorem will, for example, hold if the processes 
\begin{equation*}
\big\{H_k^{(0)},k\in\Integers^+\big\},\ldots,\big\{H_k^{(\const{L})},k\in\Integers^+\big\}
\end{equation*}
are jointly ergodic, see
\cite[Thm.~2]{kim08}.

We define the capacity \emph{pre-loglog} as
\begin{equation}
  \Lambda \triangleq \varlimsup_{\SNR\to\infty} \frac{C(\SNR)}{\log\log\SNR}.
\end{equation}
For flat-fading channels (i.e., when $\const{L}=0$) we have $\Lambda=1$
\cite{lapidothmoser03_3}. For multipath fading
channels with an infinite number of paths (i.e., when $\const{L}=\infty$),
 it has been shown in \cite{kochlapidoth08_1} that when the sequence
 $\{\alpha_{\ell}\}$  decays not faster than geometrically, then
 capacity is bounded in the SNR and hence $\Lambda=0$. One might
therefore expect that the pre-loglog decays with
$\const{L}$. It turns out, however, that this is not the
case.

\subsection{Main Result}
\label{sub:result}

\begin{theorem}
  \label{thm:main}
  Consider the above channel model, and assume that $\const{L}<\infty$. Then, irrespective of $\const{L}$, the
  capacity pre-loglog is given by
  \begin{equation}
    \Lambda = \lim_{\SNR\to\infty}\frac{C(\SNR)}{\log\log\SNR} = 1.
  \end{equation}
\end{theorem}
\begin{IEEEproof}
  See Section~\ref{sec:proof}.
\end{IEEEproof}
Thus for finite $\const{L}$, the pre-loglog is not affected by the multipath
behavior.

\section{Proof of Theorem~\ref{thm:main}}
\label{sec:proof}
In Section~\ref{sub:upper} we derive a capacity upper bound and show
that the ratio of this bound to $\log\log\SNR$ tends to $1$ as $\SNR$
tends to infinity. In Section~\ref{sub:lower} we propose a
coding scheme which achieves a capacity pre-loglog of $1$. Both
results combine to prove Theorem~\ref{thm:main}.

\subsection{Converse}
\label{sub:upper}
We begin with the chain rule for mutual information
\cite{coverthomas91}
\begin{equation}
  I\big(X_1^n;Y_1^n\big) = \sum_{k=1}^n I\big(X_1^n;Y_k\big|Y_1^{k-1}\big)\label{eq:upperchain}
\end{equation}
and upper bound then each summand on the right-hand side (RHS) of
\eqref{eq:upperchain} using the general upper bound for mutual
information \cite[Eq.~(27)]{lapidothmoser03_3}
\begin{IEEEeqnarray}{lCl}
  I\big(X_1^n;Y_k\big|Y_1^{k-1}\big)& \leq & \E{\log
    |Y_k|^2}-h\big(Y_k\big|X_1^n,Y_1^{k-1}\big)\nonumber\\
  & & {} + \xi
  \left(1+\log\E{|Y_k|^2}-\E{\log|Y_k|^2}\right)\nonumber\\
  & & {} +\log\Gamma(\xi)-\xi\log\xi+\log\pi \label{eq:duality}
\end{IEEEeqnarray}
for any $\xi>0$. Here $\Gamma(\cdot)$ denotes the Gamma
function. 

We evaluate the terms on the RHS of \eqref{eq:duality} individually.
We use \cite[Eq.~(15)]{kochlapidoth08_1} to upper bound
\begin{equation}
  \E{\log|Y_k|^2} \leq \E{\log\left(\sigma^2+\sum_{\ell=0}^{k-1}\alpha_{\ell}|X_{k-\ell}|^2\right)}\label{eq:upper1}
\end{equation}
and \cite[Eq.~(24)]{kochlapidoth08_1} to lower bound
\begin{IEEEeqnarray}{lCl}
  h\big(Y_k\big|X_1^n,Y_1^{k-1}\big) & \geq &
  \E{\log\left(\sigma^2+\sum_{\ell=0}^{k-1}\alpha_{\ell}|X_{k-\ell}|^2\right)}
  \nonumber\\
  & & {} + \inf_{\ell\in\set{L}}\left(h_{\ell}-\alpha_{\ell}\right),\label{eq:upper2}
\end{IEEEeqnarray}
where we define $\alpha_{\ell}\triangleq 0$, $\ell=\const{L}+1,\const{L}+2,\ldots$. The
next term is readily evaluated as
\begin{equation}
  \log \E{|Y_k|^2} = \log\left(\sigma^2+\sum_{\ell=0}^{k-1}\alpha_{\ell}\E{|X_{k-\ell}|^2}\right).\label{eq:upper3}
\end{equation}
Finally, we use \cite[Eq.~(30)]{kochlapidoth08_1} to lower bound
\begin{IEEEeqnarray}{lCl}
  \E{\log|Y_k|^2} & \geq &
  \E{\log\left(\sigma^2+\sum_{\ell=0}^{k-1}\alpha_{\ell}|X_{k-\ell}|^2\right)}+\log\delta^2\:\:\:\:\:\:\:\:\nonumber\\
  \IEEEeqnarraymulticol{3}{r}{{} - 2 \eps(\delta,\eta)-\frac{2}{\eta}\left(\frac{2}{e}+\log(\pi
    e)\right)+\frac{2}{\eta}
  \inf_{\ell\in\set{L}}\left(h_{\ell}-\alpha_{\ell}\right),\:\:\:\:\:\:\:\:}\label{eq:upper4}
\end{IEEEeqnarray}
where $0<\delta\leq 1$, $0<\eta<1$, and where $\eps(\delta,\eta)>0$
tends to zero as $\delta\downarrow 0$.

Subtracting \eqref{eq:upper4} from \eqref{eq:upper3}, and lower
bounding
$\E{\log\left(\sigma^2+\sum_{\ell=0}^{k-1}\alpha_{\ell}|X_{k-\ell}|^2\right)}\geq\log\sigma^2$
yields
\begin{IEEEeqnarray}{lCl}
  \IEEEeqnarraymulticol{3}{l}{\log\E{|Y_k|^2}-\E{\log|Y_k|^2}}\nonumber\\
  \quad & \leq &
  \log\left(1+\sum_{\ell=0}^{k-1}\alpha_{\ell}
  \E{|X_{k-\ell}|^2}/\sigma^2\right) + \Psi, \label{eq:upper5}
\end{IEEEeqnarray}
where we define
\begin{equation}
  \Psi \triangleq
  \log\frac{1}{\delta^2}+2\eps(\delta,\eta)+\frac{2}{\eta}\left(\frac{2}{e}+\log(\pi
  e)\right)-\frac{2}{\eta}\inf_{\ell\in\set{L}}\left(h_{\ell}-\alpha_{\ell}\right).
\end{equation}
Thus we obtain from \eqref{eq:upper5}, \eqref{eq:upper2}, \eqref{eq:upper1}, and
\eqref{eq:duality}
\begin{IEEEeqnarray}{lCl}
  \IEEEeqnarraymulticol{3}{l}{I\big(X_1^n;Y_k\big|Y_1^{k-1}\big)}\nonumber\\
  \quad & \leq &
  -\inf_{\ell\in\set{L}}\left(h_{\ell}-\alpha_{\ell}\right)\nonumber\\
  & & {} + \xi\left(
    1+\log\left(1+\sum_{\ell=0}^{k-1}\alpha_{\ell}\E{|X_{k-\ell}|^2}/\sigma^2\right)+\Psi\right) \;\;\;\nonumber\\
  & & {} + \log\Gamma(\xi)-\xi\log\xi+\log\pi.\label{eq:upper6}
\end{IEEEeqnarray}

Let $\alpha^{(\const{L})}$ be defined as
\begin{equation}
  \alpha^{(\const{L})} \triangleq \sum_{\ell=0}^\const{L} \alpha_{\ell}.
\end{equation}
We choose now
\begin{equation*}
  \xi = \left(1+\log\left(1+\alpha^{(\const{L})} \SNR\right)\right)^{-1}
\end{equation*}
and use that \cite[Eq.~(337)]{lapidothmoser03_3}
\begin{equation*}
  \log\Gamma(\xi) = \log\frac{1}{\xi} + o(1)
\end{equation*}
and that $\xi\log\xi=o(1)$
(where the term $o(1)$ vanishes as $\xi$ tends to zero) to obtain
\begin{IEEEeqnarray}{lCl}
  \IEEEeqnarraymulticol{3}{l}{I\big(X_1^n;Y_k\big|Y_1^{k-1}\big)}\nonumber\\
  \quad & \leq & -\inf_{\ell\in\set{L}} \left(h_{\ell}-\alpha_{\ell}\right)\nonumber\\
  & & {} + \frac{1+\log\left(1+\sum_{\ell=0}^{k-1}\alpha_{\ell}\E{|X_{k-\ell}|^2}/\sigma^2\right)+\Psi}{1+\log\left(1+\alpha^{(\const{L})} \SNR\right)}\nonumber\\
  & & {} + \log\left(1+\log\left(1+\alpha^{(\const{L})}
      \SNR\right)\right) + \log\pi + o(1). \IEEEeqnarraynumspace\label{eq:uppersummand}
\end{IEEEeqnarray}

Using \eqref{eq:uppersummand} in \eqref{eq:upperchain}, and noting that
$\xi$---and hence also the correction term $o(1)$---does not depend on $k$ yields then
\begin{equation}
  \frac{1}{n} I\big(X_1^n;Y_1^n\big) \leq  \log\left(1+\log\left(1+\alpha^{(\const{L})}
      \SNR\right)\right) + \Upsilon_{n,\const{P}} + o(1), \label{eq:UPPER}
\end{equation}
where we define $\Upsilon_{n,\const{P}}$ as
\begin{IEEEeqnarray}{lCl}
  \Upsilon_{n,\const{P}} & \triangleq & \frac{1+\frac{1}{n}\sum_{k=1}^n
    \log\!\left(1\!+\sum_{\ell=0}^{k-1}\alpha_{\ell}\E{|X_{k-\ell}|^2}\!/\sigma^2\right)+\Psi}{1+\log\left(1+\alpha^{(\const{L})}
      \SNR\right)}\nonumber\\
  & & {} -\inf_{\ell\in\set{L}}\left(h_{\ell}-\alpha_{\ell}\right)+\log\pi.
\end{IEEEeqnarray}
Note that by Jensen's inequality
\begin{IEEEeqnarray}{lCl}
  \IEEEeqnarraymulticol{3}{l}{\frac{1}{n}\sum_{k=1}^n
    \log\left(1+\sum_{\ell=0}^{k-1}\alpha_{\ell}\E{|X_{k-\ell}|^2}/\sigma^2\right)}\nonumber\\
  \qquad & \leq &
  \log\left(1+\frac{1}{n}\sum_{k=1}^n\sum_{\ell=0}^{k-1}\alpha_{\ell}\E{|X_{k-\ell}|^2}/\sigma^2\right)\nonumber\\
  & \leq & \log\left(1+\alpha^{(\const{L})} \SNR\right),
\end{IEEEeqnarray}
where the last inequality follows by rewriting the double sum as
$\frac{1}{n}\sum_{k=1}^n\E{|X_{k}|^2}/\sigma^2
\sum_{\ell=0}^{n-k}\alpha_{\ell}$, and by upper bounding then
$\sum_{\ell=0}^{k-n}\alpha_{\ell}\leq \alpha^{(\const{L})}$ and using the
power constraint \eqref{eq:power}. Consequently, we can upper bound
$\Upsilon_{n,\const{P}}$ by
\begin{equation}
  \Upsilon_{n,\const{P}} \leq 1+\Psi
  -\inf_{\ell\in\set{L}}\left(h_{\ell}-\alpha_{\ell}\right)+\log\pi \label{eq:upperjensen}
\end{equation}
and obtain therefore from \eqref{eq:capacity}, \eqref{eq:UPPER}, and
\eqref{eq:upperjensen}
\begin{IEEEeqnarray}{lCl}
  C(\SNR) & \leq &
  \log\left(1+\log\left(1+\alpha^{(\const{L})}\SNR\right)\right) +
  1+\Psi\nonumber\\
  & & {}
  - \inf_{\ell\in\set{L}}\left(h_{\ell}-\alpha_{\ell}\right)+\log\pi + o(1).
\end{IEEEeqnarray}
Noting that $\xi
\downarrow 0$ as $\SNR$ tends to infinity (and hence $\lim_{\SNR\to\infty}o(1)=0$), this yields the desired result
\begin{equation}
  \Lambda \triangleq \varlimsup_{\SNR\to\infty}
  \frac{C(\SNR)}{\log\log\SNR} \leq 1.
\end{equation}

\subsection{Direct Part}
\label{sub:lower}
In order to show that
\begin{equation}
  \Lambda \triangleq \varlimsup_{\SNR \to \infty}
  \frac{C(\SNR)}{\log\log\SNR} \geq \varliminf_{\SNR\to\infty}\frac{C(\SNR)}{\log\log\SNR}\geq 1 \label{eq:lowerclaim}
\end{equation}
we shall derive a capacity lower bound and analyze then its
ratio to $\log\log\SNR$ as $\SNR$ tends to infinity.

To this end we evaluate $\frac{1}{n}I(X_1^n;Y_1^n)$ for the following
distribution on the inputs $\{X_k\}$. Let
$\vect{X}_b=\big(X_{b(\const{L}+\tau)+1},\ldots,X_{(b+1)(\const{L}+\tau)}\big)$ for
some $\tau\in\Integers^+$. We
shall choose $\{\vect{X}_b\}$ to be IID with
\begin{equation*}
  \vect{X}_b = \big(\underbrace{0,\ldots,0}_{\const{L}},\tilde{X}_{b\tau+1},\ldots,\tilde{X}_{(b+1)\tau}\big),
\end{equation*}
where $\tilde{X}_{b\tau+1},\ldots,\tilde{X}_{(b+1)\tau}$ is a sequence
of independent, zero-mean, circularly-symmetric
random variables with $\log|\tilde{X}_{b\tau+\nu}|^2$ being uniformly distributed over the
interval $[\log x^2_{\textnormal{min},\nu},\log
x^2_{\textnormal{max},\nu}]$, i.e.,
\begin{equation*}
  \log |\tilde{X}_{b\tau+\nu}|^2 \sim \Uniform{[\log x^2_{\textnormal{min},\nu},\log
x^2_{\textnormal{max},\nu}]}, \quad \nu=1,\ldots,\tau.
\end{equation*}
The parameters $x_{\textnormal{min},\nu}$ and
$x_{\textnormal{max},\nu}$ will be chosen later.

Let $\kappa\triangleq\lfloor \frac{n}{\const{L}+\tau}\rfloor$, and let
$\vect{Y}_{b}$ denote the vector
$\big(Y_{b(\const{L}+\tau)+1},\ldots,Y_{(b+1)(\const{L}+\tau)}\big)$. We have
\begin{IEEEeqnarray}{lCl}
  I\big(X_1^n;Y_1^n\big) & \geq &
  I\big(\vect{X}_0^{\kappa-1};\vect{Y}_0^{\kappa-1}\big)\nonumber\\
  & = & \sum_{b=0}^{\kappa-1}
  I\big(\vect{X}_b;\vect{Y}_0^{\kappa-1}\big|\vect{X}_0^{b-1}\big)
  \nonumber\\
  & \geq & \sum_{b=0}^{\kappa-1} I(\vect{X}_b;\vect{Y}_b), \label{eq:lower1}
\end{IEEEeqnarray}
where the first inequality follows by restricting the number of
observables in each of the terms; the subsequent equality follows by
the chain rule for mutual information; and the last inequality follows by
restricting the number of observables and by noting that
$\{\vect{X}_b\}$ is IID.

We continue by lower bounding each summand on the RHS of
\eqref{eq:lower1} according to \cite[Sec.~III-B]{lapidoth05_2}. We use again the
chain rule and that reducing observations cannot increase mutual
information to obtain
\begin{IEEEeqnarray}{lCl}
  I(\vect{X}_b;\vect{Y}_b) & \geq &
  I\left(\tilde{X}_{b\tau+1}^{(b+1)\tau};Y_{b(\const{L}+\tau)+\const{L}+1}^{(b+1)(\const{L}+\tau)}\right)\nonumber\\
  & = & \sum_{\nu=1}^{\tau}
  I\left(\left.\tilde{X}_{b\tau+\nu};Y_{b(\const{L}+\tau)+\const{L}+1}^{(b+1)(\const{L}+\tau)}\right|\tilde{X}_{b\tau+1}^{b\tau+\nu-1}\right)\nonumber\\
  & \geq & \sum_{\nu=1}^{\tau}
  I(\tilde{X}_{b\tau+\nu};Y_{b(\const{L}+\tau)+\const{L}+\nu}), \label{eq:lower2}
\end{IEEEeqnarray}
where we additionally have used in the last inequality that
$\tilde{X}_{b\tau+1},\ldots,\tilde{X}_{(b+1)\tau}$ are independent.

Defining
\begin{equation}
  W_{b\tau+\nu} \triangleq \sum_{\ell=1}^\const{L}
  H_{b(\const{L}+\tau)+\const{L}+\nu}^{(\ell)} X_{b(\const{L}+\tau)+\const{L}+\nu-\ell} + Z_{b(\const{L}+\tau)+\const{L}+\nu}
\end{equation}
each summand on the RHS of \eqref{eq:lower2} can be written as
\begin{IEEEeqnarray}{lCl}
  \IEEEeqnarraymulticol{3}{l}{I(\tilde{X}_{b\tau+\nu};Y_{b(\const{L}+\tau)+\const{L}+\nu})}\nonumber\\
  \qquad \quad & = & 
  I\big(\tilde{X}_{b\tau+\nu};H_{b(\const{L}+\tau)+\const{L}+\nu}^{(0)}\tilde{X}_{b\tau+\nu}+W_{b\tau+\nu}\big).
  \IEEEeqnarraynumspace \label{eq:lower3}
\end{IEEEeqnarray}
A lower bound on \eqref{eq:lower3} follows from the following
lemma.

\begin{lemma}
  Let the random variables $X$, $H$, and $W$ have finite second
  moments. Assume that both $X$ and $H$ are of finite differential
  entropy. Finally, assume that $X$ is independent of $H$; that $X$ is
  independent of $W$; and that $X-H-W$ forms a Markov chain. Then,
  \begin{IEEEeqnarray}{lCl}
    I(X;HX+W) & \geq & h(X) - \E{\log|X|^2}+\E{\log|H|^2}\nonumber\\
    & & {} - \E{\log\left(\pi e\left(\sigma_H+\frac{\sigma_W}{|X|}\right)^2\right)},\IEEEeqnarraynumspace
  \end{IEEEeqnarray}
  where $\sigma_W^2\geq 0$ and $\sigma_H^2>0$ are the variances of $W$
  and $H$, respectively.\footnote{Note that the assumptions that $X$
  and $H$ have finite second moments and are of finite differential
  entropy guarantee that $\E{\log|X|^2}$ and $\E{\log|H|^2}$ are finite, see
  \cite[Lemma 6.7e)]{lapidothmoser03_3}.}
\end{lemma}
\begin{IEEEproof}
  See \cite[Lemma 4]{lapidoth05_2}.
\end{IEEEproof}

It can be easily verified that for the channel model given in
Section~\ref{sub:channel} and for the above coding scheme the lemma's
conditions are satisfied. We can therefore lower bound
$I(\tilde{X}_{b\tau+\nu};Y_{b(\const{L}+\tau)+\const{L}+\nu})$ by
\begin{IEEEeqnarray}{lCl}
  \IEEEeqnarraymulticol{3}{l}{I(\tilde{X}_{b\tau+\nu};Y_{b(\const{L}+\tau)+\const{L}+\nu})}\nonumber\\
  \: & \geq & h(\tilde{X}_{b\tau+\nu}) - \E{\log|\tilde{X}_{b\tau+\nu}|^2}+\E{\log\big|H_{b(\const{L}+\tau)+\const{L}+\nu}^{(0)}\big|^2}\nonumber\\
    & & {} - \E{\log\left(\pi e\left(\sqrt{\alpha_0}+\frac{\sqrt{\E{|W_{b\tau+\nu}|^2}}}{|\tilde{X}_{b\tau+\nu}|}\right)^2\right)}.\label{eq:lower4}
\end{IEEEeqnarray}
Using that the differential entropy of a circularly-symmetric random
variable is given by (e.g., \cite[Eqs.~(320) \& (316)]{lapidothmoser03_3})
\begin{equation}
  h(\tilde{X}_{b\tau+\nu}) = \E{\log|\tilde{X}_{b\tau+\nu}|^2} +
  h\big(\log|\tilde{X}_{b\tau+\nu}|^2\big) +\log\pi,
\end{equation}
and evaluating $h(\log|\tilde{X}_{b\tau+\nu}|^2)$ for our choice of
$\tilde{X}_{b\tau+\nu}$, we obtain for the first two terms on the RHS of
\eqref{eq:lower4}
\begin{equation}
  h\big(\log|\tilde{X}_{b\tau+\nu}|^2\big) - \E{\log|\tilde{X}_{b\tau+\nu}|^2}
  =
  \log\log\frac{x_{\textnormal{max},\nu}^2}{x_{\textnormal{min},\nu}^2}
  +\log\pi. \label{eq:lowerpart1}
\end{equation}
Upper bounding
\begin{IEEEeqnarray}{lCl}
  \E{|W_{b\tau+\nu}|^2} & = & \sum_{\ell=1}^\const{L} \alpha_{\ell}
  \E{|X_{b(\const{L}+\tau)+\const{L}+\nu-\ell}|^2}+\sigma^2\nonumber\\
  & \leq & \max_{\ell=0,\ldots,\nu-1} x_{\textnormal{max},\ell}^2
  \cdot \alpha^{(\const{L})}+\sigma^2
\end{IEEEeqnarray}
(where we define $x_{\textnormal{max},0}^2\triangleq 0$), and lower bounding $|\tilde{X}_{b\tau+\nu}|^2\geq
x_{\textnormal{min},\nu}^2$, the last term on the
RHS of \eqref{eq:lower4} can be upper bounded by
\begin{IEEEeqnarray}{lCl}
  \IEEEeqnarraymulticol{3}{l}{\E{\log\left(\pi
        e\left(\sqrt{\alpha_0}+\frac{\sqrt{\E{|W_{b\tau+\nu}|^2}}}{|\tilde{X}_{b\tau+\nu}|}\right)^2\right)}}\nonumber\\
  & \leq & \log\left(\!\pi e\!\left(\!\sqrt{\alpha_0}+\sqrt{\frac{\displaystyle
        \max_{\ell=0,\ldots,\nu-1} x_{\textnormal{max},\ell}^2\cdot
        \alpha^{(\const{L})}+\sigma^2}{x_{\textnormal{min},\nu}^2}}\right)^{\!2}\right)
        \!\IEEEeqnarraynumspace
        \label{eq:lowerpart2}
\end{IEEEeqnarray}
and we thus obtain from \eqref{eq:lower4}, \eqref{eq:lowerpart1}, and \eqref{eq:lowerpart2}
\begin{IEEEeqnarray}{lCl}
  \IEEEeqnarraymulticol{3}{l}{I(\tilde{X}_{b\tau+\nu};Y_{b(\const{L}+\tau)+\const{L}+\nu})}\nonumber\\
   \; & \geq &
  \log\log\frac{x_{\textnormal{max},\nu}^2}{x_{\textnormal{min},\nu}^2}
  + \E{\log\big|H_{b(\const{L}+\tau)+\const{L}+\nu}^{(0)}\big|^2} - 1 \nonumber\\
  & & {}-2 \log\left(\sqrt{\alpha_0}+\sqrt{\frac{\displaystyle
        \max_{\ell=0,\ldots,\nu-1} x_{\textnormal{max},\ell}^2\cdot
        \alpha^{(\const{L})}+\sigma^2}{x_{\textnormal{min},\nu}^2}}\right).\;\IEEEeqnarraynumspace\label{eq:lower5}
\end{IEEEeqnarray}

Following \cite[Eqs.~(102) \& (103)]{lapidoth05_2}, we choose now
(assuming that $\const{P}>1$)
\begin{IEEEeqnarray*}{lCll}
  x_{\textnormal{max},\nu}^2 & = & \const{P}^{\nu/\tau}, &
  \nu=1,\ldots,\tau\\
  x_{\textnormal{min},\nu}^2 & = &
  \const{P}^{(\nu-1)/\tau}\log\const{P}, \qquad & \nu=1,\ldots,\tau.
\end{IEEEeqnarray*}
With this choice we have
\begin{equation}
  \frac{x_{\textnormal{max},\nu}^2}{x_{\textnormal{min},\nu}^2} =  
  \frac{\const{P}^{1/\tau}}{\log\const{P}}, \qquad \nu=1,\ldots,\tau
  \label{eq:lowerchoice1}
\end{equation}
and
\begin{IEEEeqnarray}{c}
  \frac{\displaystyle \max_{\ell=0,\ldots,\nu-1}
  x_{\textnormal{max},\ell}^2}{x_{\textnormal{min},\nu}^2} = 
  \left\{\begin{array}{ll} \displaystyle 0,
      & \displaystyle \nu=1\\ \displaystyle 
      1/\log\const{P}, \quad & \displaystyle \nu=2,\ldots,\tau.
\end{array} \right.\IEEEeqnarraynumspace\label{eq:lowerchoice2}
\end{IEEEeqnarray}
Thus applying \eqref{eq:lowerchoice1} \& \eqref{eq:lowerchoice2} to
\eqref{eq:lower5} yields
\begin{IEEEeqnarray}{lCl}
  \IEEEeqnarraymulticol{3}{l}{I(\tilde{X}_{b\tau+\nu};Y_{b(\const{L}+\tau)+\const{L}+\nu})}\nonumber\\
  \quad & \geq & \log\log\frac{\const{P}^{1/\tau}}{\log\const{P}}+
  \E{\log\big|H_{b(\const{L}+\tau)+\const{L}+\nu}^{(0)}\big|^2} -1 \nonumber\\
  & & {} -
  2 \log\left(\sqrt{\alpha_0}+\sqrt{\frac{\alpha^{(\const{L})}}{\log\const{P}}+\frac{\sigma^2}{\const{P}^{(\nu-1)/\tau}\log\const{P}}}\right)
  \nonumber\\
  & \geq & \log\log\frac{\const{P}^{1/\tau}}{\log\const{P}}+
  \E{\log\big|H_{1}^{(0)}\big|^2} -1\nonumber\\
  & & {}
  -2 \log\left(\sqrt{\alpha_0}+\sqrt{\frac{\alpha^{(\const{L})}+\sigma^2}{\log\const{P}}}\right),
  \qquad \const{P} > 1,\IEEEeqnarraynumspace \label{eq:lower6}
\end{IEEEeqnarray}
where the last inequality follows by using that the process $\big\{H_{k}^{(0)},k\in\Integers^+\big\}$
is stationary and that, for $\const{P} > 1$,
\mbox{$\const{P}^{(\nu-1)/\tau}\geq 1$}.

Note that the RHS of \eqref{eq:lower6}
depends neither on $\nu$ nor on $b$. We therefore obtain from
\eqref{eq:lower6}, \eqref{eq:lower2}, and \eqref{eq:lower1}
\begin{equation}
  I\big(X_1^n;Y_1^n\big) \geq \kappa\tau
  \log\log\frac{\const{P}^{1/\tau}}{\log\const{P}} + \kappa\tau
  \Xi_{\const{P}}, \quad \const{P}> 1, \label{eq:lower7}
\end{equation}
where we define $\Xi_{\const{P}}$ as
\begin{equation}
  \Xi_{\const{P}} \triangleq \E{\log\big|H_{1}^{(0)}\big|^2} - 1  -2\log\!\left(\sqrt{\alpha_0}+\sqrt{\frac{\alpha^{(\const{L})}+\sigma^2}{\log\const{P}}}\right).\;\;\label{eq:lowerXi}
\end{equation}
Dividing the RHS of \eqref{eq:lower7} by $n$, and computing the limit as $n$ tends
to infinity yields the capacity lower bound
\begin{IEEEeqnarray}{lCl}
  C(\SNR) & \geq & \varliminf_{n\to\infty}\frac{1}{n}
  I\big(X_1^n;Y_1^n\big)\nonumber\\
  & \geq &
  \frac{\tau}{\const{L}+\tau}\log\log\frac{\const{P}^{1/\tau}}{\log\const{P}}
  + \frac{\tau}{\const{L}+\tau}\Xi_{\const{P}}, \quad \const{P}> 1, \IEEEeqnarraynumspace\label{eq:LOWER}
\end{IEEEeqnarray}
where we have used that $\lim_{n\to\infty}\kappa/n=1/(\const{L}+\tau)$.

By noting that (for any fixed $\tau$)
\begin{IEEEeqnarray}{rCl}
  \lim_{\SNR \to \infty}
  \frac{\log\log\left(\const{P}^{1/\tau}/\log\const{P}\right)}{\log\log\SNR}
  & = & 1\\
  \lim_{\SNR\to\infty} \frac{\Xi_{\const{P}}}{\log\log\SNR} & = & 0
\end{IEEEeqnarray}
we infer from \eqref{eq:LOWER} that the capacity pre-loglog $\Lambda$
is lower bounded by
\begin{equation}
  \Lambda \triangleq
  \varlimsup_{\SNR\to\infty}\frac{C(\SNR)}{\log\log\SNR} \geq
  \varliminf_{\SNR\to\infty} \frac{C(\SNR)}{\log\log\SNR} \geq
  \frac{\tau}{\const{L}+\tau}.\;\;
\end{equation}
The claim \eqref{eq:lowerclaim} follows now by
letting $\tau$ tend to infinity.

\newpage

\section{Conclusion}
\label{sec:conclusion}
We considered a discrete-time, noncoherent, multipath fading channel where
the number of paths is finite, i.e., where the channel output
is influenced by the present and by the $\const{L}$ previous channel inputs. It was
shown that, irrespective of the number of paths, the capacity
pre-loglog is $1$ (which is equal to the pre-loglog of a
flat-fading channel). Thus, when the number of paths is finite, then
the multipath behavior has no significant effect on the high-SNR
capacity. This is perhaps surprising as it has been shown in
\cite{kochlapidoth08_1} that if
the channel output is influenced by the present and by \emph{all}
previous channel inputs, and if the variances of the path gains do not decay faster than
geometrically, then capacity is bounded in the SNR. For
such channels the capacity does not tend to infinity as the SNR tends
to infinity and hence the capacity pre-loglog is zero.

The above results indicate that the high-SNR behavior of the capacity of
noncoherent multipath fading channels depends critically on the
assumed channel model. Thus, when studying such channels at high SNR,
one has to attach great importance to the channel modeling, as slight
changes in the model might lead to completely different capacity results.

\section*{Acknowledgment}
Fruitful discussions with Helmut B\"olcskei and Giuseppe Durisi are
gratefully acknowledged.


\end{document}